\documentclass[twocolumn,showpacs,prl]{revtex4}
 \usepackage{graphicx}

\begin{document}
\title{Goldstone mode kink-solitons in double layer quantum Hall systems}

\author{\thanks{e-mail: khomeriki@hotmail.com} Ramaz Khomeriki$^1$}
\author{Kieran Mullen$^{2}$}
\author{Shota revishvili$^1$}

\affiliation{$^1$Tbilisi State University, 3 Chavchavadze, Tbilisi 380028, Republic of Georgia}
\affiliation{$^{2}$University of Oklahoma, Department of Physics and Astronomy,
440 W. Brooks, 73019 OK}

\begin{abstract}
It is shown that in charge unbalanced double layer quantum Hall system with zero tunneling pseudospin Goldstone mode excitations form moving kink-soliton in weakly nonlinear limit. This charge-density localization moves with a velocity of gapless linear spin-wave mode and could be easily observed experimentally. We  claim that mentioned Goldstone mode kink-solitons define diffusionless charge transport properties in double layer quantum Hall systems.
\end{abstract}

\maketitle

\begin{figure}[t] 
\begin{center}\leavevmode
\includegraphics[angle=-90,width=1.2\linewidth]{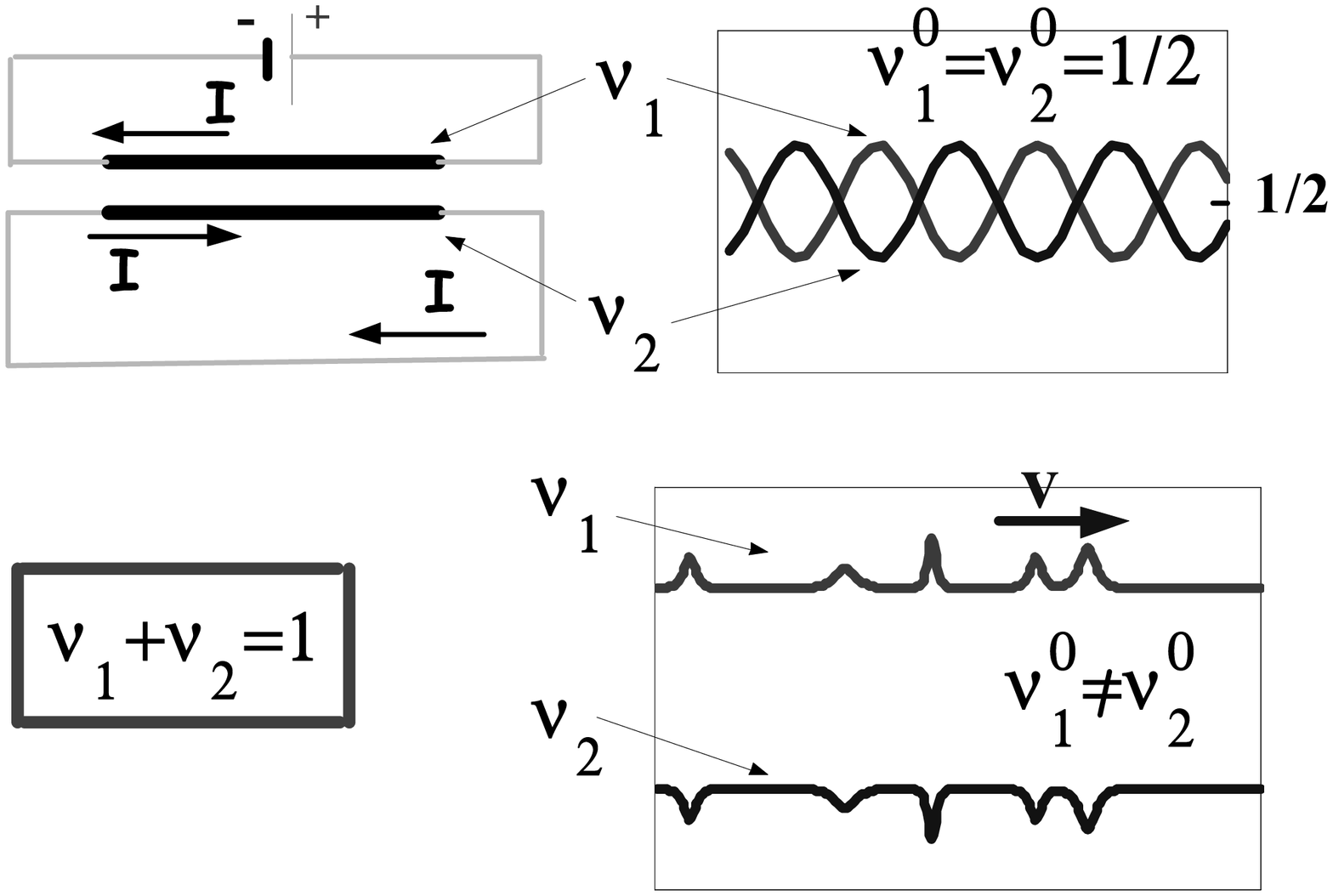}
\caption{Conditions for appearance of  linear (upper graph) and nonlinear (lower panel) pseudospin Goldstone mode excitations. In unbalanced situation when the filling factors are different from each other the nonlinear solution could be detected as counter propagating inhomogeneous currents through the layers.}
\label{lt23}\end{center}\end{figure}

The recent observation \cite{spielman2} of split-off peaks of tunneling conductance versus in-plane magnetic field  in double layer quantum Hall systems with total Landau level filling factor $\nu=1$ have been naturally interpreted as "footsteps" of linearly dispersing pseudospin Goldstone mode \cite{fertig}.
Thus the correctness and efficiency of field-theoretical approach
according which the quantum Hall double layer system could be described as easy-plane type (pseudo)ferromagnet \cite{moon} was indirectly confirmed. The Goldstone mode appears in  
systems with continuous symmetries as a result of spontaneous symmetry breaking
which in presence of nonlinearity leads to the appearance of
Goldstone mode kink-solitons \cite{hori}.
The present paper is devoted to the investigation of dynamical nonlinear properties of the mentioned Goldstone mode in quantum Hall (pseudo)ferromagnets Particularly, the existence of moving localized solutions is predicted. As it is mentioned below such localizations could be detected experimentally via observing counter propagating inhomogeneous currents through the layers (see Fig. \ref{lt23}). 

The phenomelogical model for isolated double layer quantum Hall (pseudo)ferromagnet
in the absence of tunneling is effectively described by the following Hamiltonian 
\cite{moon}:
\begin{equation}
{\mathcal H}=\frac{\rho_E}{2}\left(
\frac{\partial \bf n}{\partial x}\right)^2
+\frac{\rho_A-\rho_E}{2}\left(
\frac{\partial n_z}{\partial x}\right)^2
+\beta (n_z)^2, \label{1}
\end{equation}
where ${\bf n}(x,t)$ is an order parameter unit vector, particularly, $n_z(x,t)$ has a meaning
of local charge imbalance between the layers $n_z=\nu_1-\nu_2$ where $\nu_1$ and $\nu_2$ are filling factors for top and bottom layers, respectively (thus in fully balanced situation $\nu_1=\nu_2=1/2$); $\rho_A$ and $\rho_E$ are 
out of plane and in-plane pseudospin stiffnesses and $\beta$ gives a hard axis anisotropy.

The time-space behavior of ordering vector could be described in terms of Landau-Lifshitz equation:
\begin{equation}
\frac{\partial {\bf n}}{\partial t}={\bf n}\times{\bf
H}_{eff}, \quad
{\bf H}_{eff}=2\left\{\frac{\partial}{\partial x}\left[\frac{\partial{\mathcal H}}{\partial
\frac{\partial\bf{n}}{\partial x}}\right]-\frac{\partial{\mathcal H}}{\partial{\bf n}}\right\} \label{6}
\end{equation}
where $H_{eff}$ is the effective (pseudo)magnetic field. Then from (\ref{6}) one can write down the motion equations explicitly as
\begin{eqnarray}
\frac{\partial n^+}{\partial t}=2i\left[2\beta n^z-\rho_A
\frac{\partial^2 n^z}{\partial x^2}\right]n^++2i\rho_E
n^z\frac{\partial^2n^+}{\partial x^2} \nonumber \\ 
\frac{\partial n^z}{\partial t}=i\rho_E\frac{\partial}
{\partial x}\left[n^+\frac{\partial n^-}{\partial x}-n^-\frac{\partial n^+}
{\partial x}\right], \label{7}
\end{eqnarray}
where the following definition is used $n^\pm=n_x\pm in_y$.

The set of equations (\ref{7}) permits the uniform solution in the form:
\begin{equation}
n^+=n_\perp^0e^{4i\beta n_z^0t}; \quad n_z=n_z^0; \quad n_\perp^0=\sqrt{1-(n_z^0)^2}, 
\label{00}
\end{equation}
Preparing initially the sample with a 
given charge imbalance $n_z^0=\nu_1^0-\nu_2^0$ this quantity will stay unchanged 
for infinitely long time period as far as the isolated double layer systems are considered and tunneling does not exist. 

Let us seek for the solution of (\ref{7}) as a perturbation of uniform solution (\ref{00}):
\begin{equation}
n^+=e^{4i\beta n_z^0t}(n_\perp^0+m^+); \qquad n_z=n_z^0+m_z.  \label{100}
\end{equation}
According to the general approach \cite{taniuti} we present ${\bf m}(x,t)$ in the form of multiple scale expansion:
\begin{equation}
{\bf m}=\sum\limits_{\gamma=1}^\infty\varepsilon^\gamma 
{\bf m}^{(\gamma)}(\xi,\tau), \quad
\xi=\varepsilon (x- v t), \quad \tau=\varepsilon^3t, \label{9}
\end{equation}
where $\xi$ and $\tau$ are the slow space-time variables, $v$ is a propagation velocity of nonlinear wave
and $\varepsilon$ is a formal small parameter indicating the smallness or "slowness"
of the variable before which it appears. Considering weakly nonlinear limit (perturbative solution) we have a following restriction $({\bf m})^2\ll 1$.

Building the perturbative solution we substitute
Exps. (\ref{100}) and (\ref{9}) into the motion equation (\ref{7}) collecting the terms of the same order 
over $\varepsilon$. Using the general approach \cite{taniuti} we finally come to the Korteweg - deVries equation with a well-known one soliton solution: 
\begin{equation}
n^{+}=n_\perp^0-i\varphi; \qquad n_z=n_z^{0}+\sqrt{\frac{\rho_E}{2\beta}}\cdot\frac{\partial
\varphi}{\partial x} \label{amp}
\end{equation}
where $\varphi$ is a real function:
\begin{equation}
\varphi=A\tanh\left[\frac{x-vt}{\Lambda}\right]; \label{23}
\end{equation}
$A\ll 1$ is an amplitude of weakly nonlinear wave; $\Lambda$ is a soliton width
\begin{equation}
\Lambda=\sqrt{\frac{2\rho_E}{\beta}}\frac{(n_\perp^0)^2\rho_A+
(n_z^0)^2\rho_E}{n_z^0\rho_EA}; 
\label{24}
\end{equation}
and $v$ is a propagation velocity of the kink-soliton coinciding with the velocity of linear Goldstone mode $v=n_\perp^0\sqrt{8\beta\rho_E}$ (see e.g. Ref. \cite{moon}).

As it is seen from Eqs. (\ref{amp}) and (\ref{23}) this localized object has a kink like form
in pseudospin $xy$ plane while it is an ordinary moving soliton considering the $z$ 
component in the pseudospin space. From the same expressions we also see that in fully balanced case
$n_z^0\rightarrow 0$ the kink-soliton solution disappears - its localization width $\Lambda$ and excitation amplitude
go to infinity and zero, respectively. Thus we can conclude that for balanced situation and nonzero distance between the layers the linearized Goldstone mode is an exact solution even in the nonlinear limit while in unbalanced state $n_z^0=\nu_1^0-\nu_2^0\neq 0$ only excitations in the system are Goldstone mode kink-solitons (see Fig. \ref{lt23}).

Considered kink-solitons are moving charged localizations and therefore could be detected experimentally. Indeed, any perturbation of unbalanced double layer system will induce counter propagating inhomogeneous electric current through the layers. We can also take into account the dissipation effects adding to the Landau-Lifshitz equation (\ref{6}) a damping term. This will cause decreasing (in time) of amplitude of kink-soliton with its simultaneous widening like it happens in case of magnetization solitons in magnetic thin films \cite{ref}.  

\vspace{2cm}

{\bf Acknowledgements:}
The work at the University of Oklahoma was supported by the NSF under grant No.
EPS-9720651 and a grant from the Oklahoma State Regents for higher education.
The research described in this publication was made possible in part by Award No. GP2-2311-TB-02 of the U.S. Civilian Research \& Development Foundation for the Independent States of the Former Soviet Union (CRDF).

\end{document}